\documentclass[11pt]{article}
\usepackage{DGfest,epsfig}
\usepackage{times}
\def\be{\begin{equation}}
\def\ee{\end{equation}}
\def\bea{\begin{eqnarray}}
\def\eea{\end{eqnarray}}

\begin{document}
\baselineskip 11.5pt
\title{Theories, models, simulations: a computational challenge}

\author{G.C. ROSSI~\footnote{Talk given at the miniconference, 
``Sense of Beauty in Physics'', in honor of Adriano Di Giacomo, Pisa, January 26-27, 2006}}

\address{Dipartimento di Fisica, Universit\`a di Roma {\it Tor Vergata},\\
INFN, Sezione di Roma {\it Tor Vergata}\\
Via della Ricerca Scientifica, 00133 Roma, Italy}

\maketitle\abstracts{In this talk I would like to illustrate with examples 
taken from Quantum Field Theory and Biophysics how an intelligent 
exploitation of the unprecedented power of today's computers could led not 
only to the solution of pivotal problems in the theory of Strong 
Interactions, but also to the emergence of new lines of interdisciplinary 
research, while at the same time pushing the limits of modeling to the 
realm of living systems.}

\section*{Prologue}
\label{sec:PRO}

The somewhat schematic partition of the last century natural science 
into separated fields of research, which were essentially identified with 
mathematics, physics and biology, is nowadays becoming less and less rigid,
leading to large areas of overlapping interests.

The fundamental reason for the former {\it de facto} separation was 
the enormous amount of accumulated knowledge in each of the three areas,
which resulted in an increasing, and at the end unsurmountable, degree of 
specialization for people working at the front-end of their research field. 

Two facts have been drastically changing the situation. One was 
the growing evidence that methods and ideas developed in one 
research area could be fruitfully exported to other, even distant, 
fields of investigation. The second, is a related one and has to do with 
the sharp increase of the available computing resources (in terms 
of CPU-time, memory and storing capacity), which is making algorithms and 
general computational strategies immediately ready for use to researchers 
working in different areas. 

In my opinion this last fact is of particular relevance in 
today's spectacular progress of science, because it has 
allowed to imagine and attack problems that were considered 
impossibly difficult only a few years ago. New, flexible and adaptive 
computational tools that can be of general help to many scientific disciplines 
are being implemented under the pressure of the challenges posed on the 
one hand by the developments of pure science and technology and on the 
other by the fast expanding needs of our modern societies (think to weather 
forecasting, stock market ``surveillance'', power plant control systems, 
distributed information network management, etc.).

Taking an example of this trend from a field which is nearer to the 
scientific interests of our community, it is interesting to remark that one 
of the most extraordinary and somewhat unexpected outcome of the long lasting 
interplay between Statistical Mechanics and the theory of Strong 
Interactions in its lattice formulation (lattice QCD - LQCD) was the 
decision taken within the community of theoretical physicists to 
build ``dedicated machines'' with parallel architecture~\cite{APE,COL}.
The aim of these machines was to provide a tool capable of efficiently 
dealing with the extremely hard computational task of extracting useful 
physical information from the simulation of QCD, when the latter is seen 
as a statistical system of interacting ``coloured spins'' living on the 
sites of a (Euclidean space-time) lattice with gauge fields sitting on the links. 

Numerical and conceptual tools developed in Statistical Mechanics and in 
Theoretical Chemistry~\cite{WIL,METR} immediately found applications in 
LQCD, and vice-versa ideas and numerical techniques invented in LQCD were 
fed back in simulations of statistical systems~\cite{PAR,AMIT} 
as well as in the study of the more complicated situations that appear 
when systems of biological interest are modeled~\cite{BERG,OKA}.

The second half of the 80's was marked by a breakthrough in the theory 
of disordered systems that turned out to have a significant impact in numerous 
emerging fields of investigation. The replica approach was extended to spin glass 
systems and the notion of replica symmetry breaking was proposed as an 
explanation for the occurrence of the glassy phase transition~\cite{PMV}. 
In this context a new and more precise notion of complexity has emerged, 
suggested by the phenomenology of spin glass, that rather soon appeared 
to be of great relevance in the apparently distant problem of 
constructing mathematically sensible models of biosystems.

In fact, there is an intriguing analogy between the mathematical structure 
of spin glasses~\cite{PMV} and certain approaches to the problem of modeling protein 
folding~\cite{MAR,PJM,OKA}. Here the relation between the two fields is in physical 
terms less direct than in the case of Quantum Field Theory and Statistical 
Mechanics mentioned above and most importantly in the case of both spin glasses and proteins  
mathematical computationability is intrinsically limited by the complexity 
of the models one is considering. Despite these difficulties, a lot has been 
learned about protein structure from approaches inspired by the theory of disordered 
systems and, vice-versa, ideas taken from biology have spurred new strategies aimed at 
dealing with hard computational problems (NP-complete problems~\cite{NP,MON,ZEC})
from a novel point of view.

Indeed, the very recent discovery that the ``typical'' (not the worst) 
NP-complete problem~\cite{NP} (examples of NP-complete problems are the 
$K$-SAT problems~\cite{KSAT}) may be (almost always) solved with 
polynomial algorithms (like the cavity method, or the 
``survey inspired decimation'') seems to suggest and make us hope that 
similar methods could be developed and used to attack the most challenging 
among the theoretical problems that arise in modeling biological macro-molecule 
interactions, among which we might mention protein folding and aggregation, 
protein-protein and protein-DNA recognition, etc.

I cannot end this brief overview without recalling that the most spectacular 
and successful results of the generalized use of computers in pure 
and applied research are probably to be found in the realm of life sciences. 
Sequencing the human genome would have been impossible without the support of the 
most advanced computers of the time~\cite{SCNAT}. Today the big task is annotation. 
It is now clear that to gain a really useful understanding about the complexity 
of living systems, we need to record, cross-link and organize in an appropriate 
way the exponentially fast growing amount of biological information that is being 
gathered in experiments. The task is made particularly difficult by the impressive 
variety of data we need to store and correlate. Just to give you some examples 
of such an enormous variety let me recall that understanding biosystems at large will 
require dealing with data that go from the structure of the metabolic networks of 
biochemical reactions taking place in the cell to the description of the series 
of events by which the immunological system responds to an antigen, from the 
epidemiological and statistical information necessary to monitor the progression 
and the spreading of a disease in a population to the biochemical characterization 
of the complicated protein-DNA interactions which regulate gene expression and so on. 

The very same development of the micro-array technique, that so much 
biological information is continuously providing, was only possible thanks 
to the wide-spread availability of computers capable to deal with 
the huge outflow of data of combinatorial chemistry in an efficient, 
reliable and retrievable way.

\section{Introduction}
\label{sec:INTRO}

Personally I was introduced to the fascinating field of computers and 
simulations by Adriano in 1980, when we were both visiting CERN. It was the 
exciting time when the first attempts to extract physics from numerical 
simulations of QCD were just starting to produce useful results and APE 
was a new extraordinary scientific and technological enterprise. 

Since then the increase of the computational power at disposal to 
research and every-day life has proceeded at a pace that only 
the most blunt extrapolation of the Moore law~\footnote{Moore's original 
statement was the observation made in 1965~\cite{MOO} that the number 
of transistors per square inch on integrated circuits (we would more precisely say 
today the number of transistors that minimizes the cost per transistor in a chip) 
had doubled every year since the integrated circuit was invented. Moore, co-founder 
of Intel, predicted that this trend would continue for the foreseeable future. 
In subsequent years, the pace slowed down a bit, but transistor density has 
doubled approximately every 18 months, and this is the current definition of 
Moore's Law, which Moore himself has blessed. Most experts, including Moore 
himself, expect Moore's Law to hold for at least another two decades.} over 
more than thirty years was able to predict. This exponential explosion has 
radically changed not only the life style of billions of people, but 
also the way we scientists think about science and research. Completely 
new problems have appeared to be within our reach, that only few years ago would 
have seemed just impossible to attack or even to dream of. If appropriately used, 
computers represent more than a simple tool which can increase our 
ability to answer questions: their enormous potentiality, associated 
to flexibility and adaptability, has opened the way to new adventures 
that are only limited by our fantasy and courage.

In this talk I would like to try to underline the irreplaceable role of what 
might be called ``intelligent computing'' in certain domains of physics and 
biophysics, by illustrating in three significative examples of application, 
chosen according to my personal inclination and competences, how new ideas 
could be effectively implemented and made to work thanks to the power of 
the available computational means. Two examples are taken from the field of 
Monte Carlo simulations of LQCD. The first has to do with the analysis 
of the gluon sector of QCD (sect.~\ref{sec:FF2}). The second with possible ways 
of solving or easing the problems posed by the explicit breaking of chiral 
symmetry which according to the Nielsen--Ninomiya theorem~\cite{NN} affects any 
(ultra-)local lattice regularization of QCD (sect.~\ref{sec:FCS}). In the third 
example I wish to report on a somewhat innovative approach to the study of polymer 
structure with the methods of Statistical Mechanics (sect.~\ref{sec:SPP}).

\section{Gluon operators}
\label{sec:FF2}
 
I want to start by discussing two selected topics related to the gluonic sector 
of QCD where ``intelligent computing'' has been decisive to give support to our 
understanding of certain properties of the Theory of Strong Interactions.
I will illustrate the calculation and the physical relevance of two quantities: 
the plaquette expectation value and the topological susceptibility. The first 
quantity is related to the so-called gluon condensate~\cite{SVZ}. The second is 
supposed to be responsible~\cite{WV} for the non-vanishing of the $\eta'$ mass 
in the chiral limit (the limit where quark masses are sent to zero).

\subsection{The plaquette expectation value}
\label{sec:PLAQ}

The expectation value of the plaquette, $\langle P\rangle$, is an obviously 
relevant quantity in the study of the thermodynamic properties of lattice   
gauge theories. Besides, it was thought that one could extract the $F^2$-gluon 
condensate of ref.~\cite{SVZ} from lattice data if one could subtract from the 
lattice data on $\langle P\rangle$ its perturbative tail~\cite{DGR}. In this context 
it was an open question to decide whether signs of renormalon effects~\cite{RENO}
and of what dimension were visible in the plaquette perturbative expansion. 

At the time where we (I mean Adriano and me) started to ask ourselves such 
questions there was little experience about perturbative and non-perturbative 
definition of lattice composite operators and even less about the relation 
between lattice and continuum expectation values. Lacking any better strategy, 
we attacked by brute force the problem of defining the $F^2$-operator starting 
from its definition in terms of the plaquette expectation value. We computed 
the first three terms ({\it i.e.}\ tree-level, order $g^2$ (1-loop) and order 
$g^4$ (2-loops)) in the perturbative expansion of the plaquette by hand. At that 
time ours was the most difficult perturbative lattice calculation ever attempted. 
It took us about six months of intense work and cross-checking until we could 
agree on the analytic expression of the function that we then had to integrate 
numerically~\cite{DGR}. The result was
\begin{equation}
\langle 1-P\rangle= \frac{1}{4}\frac{N_c^2-1}{2N_c}g^2+
\frac{1}{2N_c}\frac{N_c^2-1}{2N_c}
\Big{[}(0.0203\pm 0.0001)N_c^2-\frac{1}{32}\Big{]} g^4+{\mbox{O}(g^6)}\, ,
\label{LOWEST}\end{equation}
where $N_c$ is the number of colours and the error in parenthesis comes
from the uncertainty inherent in the numerical integration. It is amazing 
(should I say disappointing looking back at our effort?) to observe that 
the clever stochastic methods which are available today~\cite{STOC} allow 
to compute the perturbative expansion of $\langle P\rangle$ up to order 
$(g^2)^{16}$, with the aid of a good 32-node cluster in a few 
hours~\cite{RAK}. From this recent knowledge indications are that a dimension 
four operator can indeed be seen below the computed perturbative tail, if 
accurate data, like those of ref.~\cite{LIMEU}, are used in the analysis.

To tell the truth we could do something slightly better: by comparing our 
results to the brand new $N_c=2$ simulation data just produced in those days 
by Mike Creutz~\cite{CRE}, we could extract the numerical value of the 
coefficient of the term $g^6$ obtaining an estimate which, within its 
relatively large error, appears to be quite accurate, when compared 
to the successive explicit perturbative calculations of ref.~\cite{ACFP}.  

This story is paradigmatic of the inextricable interplay between technological 
developments and scientific intelligence. Thanks to his scientific creativity 
Mike Creutz was able to exploit at their best the computational possibilities 
of the time, producing data that led us to ask questions whose answer had 
in turn to wait still a few years before one could arrive at the technical 
improvements and theoretical advances necessary to get a full comprehension 
of the underlying problems.

\subsection{Topological charge density and susceptibility}
\label{TCDS}

Topology is a key concept in gauge theories. According to our understanding 
of the solution of the so-called U(1)$_{\rm{A}}$ problem a non-vanishing 
topological susceptibility is responsible for providing a mass to the 
$\eta'$ pseudo-scalar meson in the limit where up, down and strange quark 
masses are set to zero. As a result the $\eta'$ is not the ninth Goldstone 
boson of chiral symmetry. 

In the limit $N_f/N_c\to 0$~\footnote{$N_f$ is 
the number of light (massless) quark flavours.} the mass of the (lightest) 
flavour singlet pseudo-scalar meson is given by the well-known Witten--Veneziano 
(WV) formula~\cite{WV}
\begin{equation}
m_{\eta '}^2=\frac{2N_f}{F_\pi^2}A\, , 
\label{MASS}\end{equation}
where $F_\pi$ is the pion decay constant (normalized so that 
$F_\pi\simeq 94\,{\rm{MeV}}$ for $N_f=3$) and $A$ is the ``topological
susceptibility". $A$ is formally defined by the equation 
\begin{equation}
A=\int d^4x\, \langle Q(x) Q(0)\rangle\Big{|}_{\rm{YM}}\, , 
\label{A}\end{equation}
with $Q(x)$ the topological charge density, which in the formal continuum theory 
has the expression 
\begin{equation}
Q(x)=\frac{g^2}{64\pi^2}\epsilon_{\mu\nu\rho\sigma} 
\sum_{a=1}^{N_c^2-1}F^a_{\mu\nu}F^a_{\rho\sigma}(x) \, .
\label{Q}\end{equation}
The notation $\langle\ldots\rangle|_{\rm{YM}}$ in eq.~(\ref{A}) means that 
the $QQ$-correlation function is to be computed in the pure Yang--Mills 
theory, {\it i.e.}\ in the absence of quarks. 

The idea that the non-perturbative value of $A$ could be measured from pure gauge 
lattice simulations dates back to the works of ref.~\cite{QTOP}, where the 
first attempts to extract such a number from numerical data were made. The 
resulting quantity, though non-vanishing and endowed with the correct scaling 
behaviour, was yielding a value of the $\eta'$ mass significantly smaller than 
phenomenologically required.  

The discrepancy was due to the fact that the renormalization effects necessary 
to match lattice and continuum definitions of topological charge density 
had been completely overlooked. This mistake was corrected in the seminal paper 
of ref.~\cite{DEF}, where the required renormalization constant was computed 
to one-loop in perturbation theory. Remarkably when the perturbatively normalized 
and vacuum subtracted simulation data for the topological susceptibility were 
inserted in eq.~(\ref{MASS}), the agreement between the theoretical calculation 
and the experimental value of the $\eta'$ mass turned out to be rather good. 

In my opinion getting an agreement between theory and experiments in this 
corner of the theory is of especially great conceptual importance, because 
the $\eta'$ mass issue is one of the few instances where the non-perturbative 
structure of QCD as a theory for Strong Interactions is at stake and can be 
subjected to a stringent test. For this very good reason the Pisa group 
(led by Adriano) has striven for some time to arrive at an accurate and 
fully non-perturbative definition of the topological objects relevant to 
this problem. Indeed they have been finally able to get a reliable 
non-perturbative determination of the renormalization constant and subtraction 
term necessary to construct from simulation data the proper definition of $A$. 
This was achieved relying on the clever method of cooling the gauge configurations 
to freeze out their perturbative fluctuations~\cite{COO}. 

\subsection{Topology in chiral regularizations of QCD}
\label{sec:TCR}

The situation of LQCD simulations has radically changed recently owing to the 
appearance on the market of exactly chiral fermions~\cite{CHI,WALL,PERF,LU} and 
the subsequent observation that the index theorem holds true as a lattice 
identity~\cite{PERF,LU} if fermions obeying the Ginsparg--Wilson (GW)~\cite{GW} 
relation are employed. In this framework the WV formula can be given a rigorous 
non-perturbative status~\cite{GRTV}. In fact, after identifying the unrenormalized 
operator which represents the topological charge density on the lattice as the one 
suggested by the flavour singlet Ward--Takahashi identities of the GW-regularized 
theory, one can prove that eqs.~(\ref{A}) and~(\ref{MASS}) are valid on the lattice 
with no need for any renormalization or subtraction.

The trouble with this approach is that simulations where the definition of $A$
suggested by GW fermions is employed are fairly expensive, although rather nice results 
have been recently obtained for it~\cite{GG}. Adriano's recent idea in this context 
is surprisingly simple and effective: it consists in making use of the GW-inspired 
definition of topological charge density only to the extent the latter is needed 
to determine the non-perturbative normalization constant of the more standard 
gluon definition~\cite{DIGNEW}, {\it i.e.}\ only to measure the topological charge 
of a configuration. The interest of this strategy is obvious: it allows to get an 
accurately normalized topological charge density without having to pay 
a much too high computational price.

\section{Waiting for a fully chiral simulation of LQCD }
\label{sec:FCS}

The next generation of computers may allow LQCD simulations with exactly chirally 
invariant fermions, {\it i.e.}\ fermions obeying the GW-condition~\cite{GW}. In 
the meantime a viable alternative~\footnote{Staggered fermions~\cite{STAG} have 
also offered a successful computational scheme.} could be to employ maximally twisted 
Wilson fermions~\cite{TM,FR1,FRC,FR2,FMPR}, possibly accompanied with a judicious 
choice of the pure gauge action. Preliminary quenched~\cite{ENC} as
well as unquenched~\cite{MONT} numerical results in this direction are quite 
encouraging. They confirm the theoretical expectation that correlators are O($a$) 
improved and that simulations require computational times that are 
of the same order of magnitude as for plain Wilson fermions (see, however, 
sect.~\ref{sec:WITC} for some word of caution). Extrapolation 
of the present trends makes us confident that the overall computational 
power allocated in Europe to maximally twisted lattice QCD (Mtm-LQCD) 
simulations can match the CPU-time needed for a study of the full theory 
in physically realistic conditions, {\it i.e.}\ on a (3~fm)$^3\times$6~fm 
lattice with a pion mass of about 250~MeV. The computation requires an estimated 
power of the order of 10 Teraflop*year~\footnote{I wish to thank I. Montvay 
for correspondence on this issue.}. Optimistically one may hope to get the 
first useful results in a little more than one year from now. In view of this 
remarkable and fortunate situation I think it might be worth reviewing the 
theoretical structure and the properties of Mtm-LQCD as developed in 
refs.~\cite{FR1,FRC,FR2,FMPR}. 

\subsection{A cheap proposal}
\label{sec:ACP}

Soon after noticing that to avoid exceptional configurations in Wilson 
fermion simulations one should introduce quarks in flavour pairs 
and have the Wilson term rotated with respect to the quark mass term 
by an axial rotation in iso-spin space, it was realized that an 
especially useful choice for that angle is to set it at its maximal
value, $|\omega|=\pi/2$, because in such a situation O($a$) (actually 
O($a^{2k+1}$), $k\geq 0$) improvement of physical quantities is 
automatic with no need to introduce the ``clover term''~\cite{SW} in the action. 


It was then shown in~\cite{FRC} that the nice improvement properties enjoyed 
by Mtm-LQCD, which were derived for pairs of mass degenerate quarks in~\cite{FR1}, 
can be immediately extended to the more interesting case of non-degenerate quarks, 
without loosing the positivity of the corresponding fermion determinant. The last 
property is obviously crucial if one wants to be able to set up workable 
Monte Carlo-like simulation algorithms for LQCD.

With the above ingredients and exploiting the flexibility offered by the freedom 
of regularizing different valence flavours with different values of the Wilson 
parameter, it was shown in~\cite{FR2} that it is possible to construct a hybrid theory, 
where sea quarks are introduced as pairs of non-degenerate particles and valence quarks 
are regularized as \"Osterwalder--Seiler~\cite{OS} fermions, such that no 
``wrong chirality'' mixing~\cite{BMMRT} affects the computation of the matrix 
elements of the ${\cal{CP}}$-conserving $\Delta S=1,2$ effective weak Hamiltonian.
Of course the same result would hold if GW fermions were used as valence quarks.
Absence of wrong chirality mixing makes Mtm-LQCD a more appealing regularization 
of QCD than the one offered by the use of standard (clover) Wilson fermions. 

{}From what we said above about improvement, it turns out that Mtm-LQCD correlators 
that are not trivially vanishing in the continuum limit can be affected by lattice 
artifacts described by a Symanzik expansion~\cite{SYMA} with only even powers of $a$.
Among these terms there are lattice contributions which tend to become large as the 
quark mass is lowered. They originate from the breaking of parity and iso-spin 
induced by the presence of the twisted Wilson term in the action. These effects  
have been discussed both in chiral perturbation theory~\cite{SHWUNEW,AB}, as well as 
in the language of the Symanzik expansion~\cite{FMPR} where they appear as terms of 
the form $(a/m_q)^{2k}$, $k\geq 1$. The general conclusion of the theoretical 
analysis is that such lattice artifacts can be reduced to a numerically tolerable 
level (precisely down to order $a^2(a^2/m_q)^{k-1}$, $k\geq 1$) if the clover 
term~\cite{SW} is introduced in the action~\cite{FMPR} (with its non-perturbatively 
determined $c_{SW}$ coefficient~\cite{LUSS}) or, alternatively, if the critical 
mass is chosen in some ``optimal way''~\cite{SHWUNEW,AB,FMPR}. 
Actually it turns out~\cite{FGR} that, at least up to O($a$) included, the optimal 
critical mass coincides with the critical mass one would get from the vanishing of 
the pion mass (or the PCAC mass) within the standard Wilson fermion regularization.

The previous discussion about chirally enhanced discretization artifacts affecting 
Mtm-LQCD correlators is rather important because it shows that the strong (order of 
magnitude) inequality
\begin{equation} m_q> a\Lambda^2_{\rm QCD}\, , \label{STINEQ}\end{equation}
invoked in ref.~\cite{FR1} in order to have the phase of the chiral 
vacuum driven by the quark mass term and not by the (twisted) Wilson 
term, can be relaxed to the more favourable relation 
\begin{equation} m_q > a^2\Lambda^3_{\rm QCD}\, , \label{WINEQ}\end{equation}
before large cutoff effects are possibly met when the quark mass is 
lowered at fixed $a$. The bound~(\ref{WINEQ}) is fairly weak as it 
permits simulations in a region of quark masses that correspond to 
rather light pions (with masses around 200~MeV for typical 
present-day lattice spacings). 

\subsection{Where is the catch?}
\label{sec:WITC}

All this sounds good, perhaps too good to be true. So the natural question 
to ask is: is there a catch in the twisted mass approach to LQCD and where is it? 

To tell the truth there is one little catch. It has to do with the 
observation~\cite{META1,META2,CPT} that at too coarse lattice spacing 
meta-stabilities are seen to affect unquenched data~\cite{MONT} which 
prevent their extrapolation to the chiral limit. Such meta-stabilities 
are the consequence of the explicit breaking of chiral symmetry induced 
by the presence of the Wilson term in the action. They appear at 
sufficiently low quark mass when the latter is progressively lowered 
at fixed $a$ and cause the statistical system one is dealing 
with not to reach equilibrium. For recent reviews on these and related 
questions and an updated assessment of the present status of quenched 
and unquenched Mtm-LQCD simulations see ref.~\cite{SHI}. 

A safe way-out of these difficulties is obviously to work at sufficiently small 
lattice spacing: something which, however, may turn out to be computationally 
too expensive. Actually there is another, more clever solution to the existence of 
meta-stable phases (and to the other dangerous flavour breaking effects~\cite{CHRMI} 
that in this regime plague the theory) which will work even on coarse lattices. 
It consists in tuning the pure gauge action so as to set to zero (in the 
chiral limit) the matrix element of the dimension six operator of the Symanzik low 
energy action of LQCD taken between pion states with vanishing three-momentum, 
{\it i.e.}\ the quantity $c_2\propto\langle\pi({\bf 0})|{\cal{L}}_6|\pi({\bf 0})\rangle$. 
It can be shown that this particular matrix elements controls the magnitude of 
all the unwanted cutoff effects described above~\cite{FRCY,SHWUNEW,AB}, 
making them to vanish as soon as $c_2=0$.

This strategy has been already partially implemented by working with a gluon 
action other that the standard plaquette action~\cite{MONT,SHI}. One finds that 
meta-stabilities are avoided in this way as soon as $a\leq 0.1$~fm and for pion 
masses down to (at least) 300~MeV.

\section{Structural properties of polymer chains}
\label{sec:SPP}

Lacking at the moment in most cases mesoscopic, functionally useful,  
descriptions of biological systems, theoretical models aimed at 
understanding the dynamic and/or the thermodynamic properties of molecular 
aggregates of biological interest are based on a detailed atomistic 
description of the compound. The physico-chemical behaviour of the 
resulting model and its compatibility with the available experimental 
information is then investigated by numerical simulations. The 
deterministic approach of Molecular Dynamics (MD) or the stochastic 
methods of Monte Carlo type~\cite{ATFR,BERG,OKA} are employed either 
classically or with quantum corrections injected {\it \'a la} 
Car-Parrinello~\cite{CP}. 

The need for a numerical approach appears to be even stronger 
if the problem of predicting the folded configuration of a protein, 
solely from the knowledge of its linear amino-acidic composition, 
is considered. The interest of investigating the folding problem 
rests on the experimental observation that the biological functionality 
of a protein crucially depends on the nature of its folded configuration~\cite{CB}. 
Misfolding is, in fact, known to lead to malfunctioning and in certain 
cases to severe pathologies, such as Creutzfeld--Jacobs disease and human 
variant of BSE~\cite{PRU}, Alzheimer disease~\cite{AD}, cystic 
fibrosis~\cite{FIB} and probably also to other neuro-degenerative processes.

Understanding the nature of folding is expected to be a formidable 
task: already the classical problem of finding the absolute minimum of 
the free energy of atomistic models of long polymer chains has a 
computational complexity which bears close resemblance to that of instances 
belonging to the class of problems technically called NP-complete~\cite{NP}.
Furthermore the problem may not have a unique solution: the recent studies on 
misfolding induced deseases have shown that proteins may live in more than one 
(meta-)stable state. It is remarkable that the simple model of ref.~\cite{PJM}  
can yield some understanding for this behaviour.

In the following sections I shall discuss merits and limitations of some interesting 
research lines and computational strategies that have been recently put forward to
deal with the problem of folding or, more modestly, with the problem of predicting 
the structure of a polymeric chain from its chemical composition. For a review 
of approaches of different nature see, for instance, ref.~\cite{MAR}.

\subsection{State of the art}
\label{sec:SOTA}

The study of Statistical Mechanics of polymers, {\it i.e.}\ long chains 
consisting of monomers of specific nature, is becoming more and more important
in chemical technologies and biological applications. Polymers like
proteins~\cite{Branden99}, nucleic acids~\cite{Saenger84},
polysaccharides~\cite{Rao98} and synthetic materials~\cite{Mattice94} 
display features that strongly depend on their detailed physico-chemical 
properties like, for instance, the degree of flexibility of certain 
chemical bonds, the charge density at monomer atoms, the structure 
of the hydrogen bond network between monomers either close or far away in
the sequence, and so on.

As we said, the enormous complications associated, even within classical
physics, with the atomistic description of the specific
interaction among the elementary components of the polymer can
only be handled by numerical simulations. Clever algorithms have
been devised to explore the configuration space
available to the system and different types of {\it ensembles}
have been invented and numerically implemented, starting from
molecular dynamics (MD) and Monte Carlo (MC) methods~\cite{ATFR}. 
As is well known, MD and MC simulations explore the {\it micro-canonical}
ensemble and the {\it canonical} ensemble of the system, respectively. 
Other kinds of ensembles, which may be collectively
indicated by the name of {\it generalized ensembles}~\cite{NOSEHOOVER},
have also been introduced and employed for the study of 
thermodynamic properties at equilibrium. In principle, under
standard ergodicity assumptions, all these {\it ensembles} should yield
equivalent physical information. Indeed the use of  
{\it generalized ensembles} within MD and MC simulation 
strategies resulted in a rather powerful approach capable of predicting the 
statistical properties of atomistic models of fluids and other compounds 
of chemical and/or biological interest in a wide range of temperatures and 
order parameter values~\cite{KLEIN}.

The crucial limitation that is encountered in numerical
simulations of systems with many relevant degrees of freedom 
is related to the inadequate and strongly biased sampling
of the configurational space occurring when the temperature is
lower than the critical temperature of the model. By ``critical
temperature'' we generically mean the temperature above which the
system is in the disordered phase. Below the critical temperature
the system remains trapped in local minima, within energy barriers
that are rarely (or never) overtaken by thermal fluctuations. 

Many strategies have been proposed over the years aimed at trying 
to overcome this difficulty. Among them we may recall simulated 
annealing~\cite{Kirkpatrick83,Okamoto01}, stochastic 
tunneling~\cite{Wenzel99} and many variants of genetic 
algorithms~\cite{Holland75}. The problem with these approaches is 
that they do not always permit the calculation of statistical averages 
in well defined {\it ensembles} ({\it i.e.}\ in the {\it ensembles} that 
are statistically representative of the desired experimental conditions). 

Vice-versa, algorithms designed to access {\it ensembles} of the 
{\it multi-canonical} type, {\it i.e.}\ the kind of {\it generalized ensembles} 
that exploit the information on the (potential) energy density of 
states~\cite{BERG,OKA}, are well assessed and made rather effective 
if used in conjunction with the replica-exchange method~\cite{MP,OKA,Okamoto01}. 
In many interesting instances it is possible to computationally monitor 
order parameters of geometrically constrained molecular models 
of polymers in a fairly large temperature range. 

However, even within these approaches problems arise when all the many degrees 
of freedom of realistic models, including the high-frequency vibration modes, are 
taken into account~\cite{MODSIM}, as it is necessary to do in order to treat 
condensed phases and explicit solvents. In fact, large variations of the 
potential energy are observed associated with such stiff terms in the Hamiltonian 
even for tiny configurational changes. Such large potential energy changes 
cause very low acceptance in the exchange of temperatures between
replicas and lack of convergence in the {\it multi-canonical} weight
computation~\cite{BERG,MODSIM}.

\subsection{A proposal for a new approach}
\label{sec:ANP}

The main lesson one learns from the previous discussion is that energy is not the 
best variable to label configurations because on the one hand configurations 
that are only slightly different in their atomic spatial arrangement may have largely 
different potential energies and on the other configurations with similar 
energy can be structurally very different. This is the main reason why, within the 
standard {\it multi-canonical} approach, introducing the temperature
through the modulation of the energy density of states by the Boltzmann 
factor does not yield sufficiently satisfactory results, as soon as the number 
of degrees of freedom of the polymer is too large and/or the 
temperature is above the order-disorder phase transition. 

In order to overcome this type of problems it was proposed in~\cite{LMPR} to work
in a {\it generalized ensemble} where configurations are generated according to 
the density of states associated to some configurational quantity (or some set of 
configurational quantities), rather than energy. These configurational quantities 
can be the mean value of some bond or dihedral angle along the polymer chain, the 
value of the $\alpha$-helicity of the polymer, the head-to-tail distance of the 
chain or any other variable which may serve to characterize the geometrical 
structure of the system.

A problem with this approach may be considered the fact that it is not clear 
how to introduce the notion of temperature, because the fundamental 
statistico-mechanical relation between the energy of the system and its temperature 
is put at stake. Actually, for the purpose of studying, say, biopolymers, which 
after all are neither isolated systems, nor do they work at equilibrium, this 
state of affairs is not really a problem and can be dealt with along the lines 
described below in sect.~\ref{sec:IT} (see also the Appendix).

In brief the idea of ref.~\cite{LMPR} is to start by working in the 
{\it micro-canonical ensemble} associated to some configurational variable, $A$, 
rather than energy, and then pass to the associated {\it canonical ensemble} by 
the ``constrained maximal entropy method'' (CMEM) imposing that $A$ assumes some 
preassigned value, $\bar a$. The latter can be either taken from experiments or can be 
known from some exact theoretical calculations in particularly simple models. The scheme
can be extended in an obvious way to the case of more than one configurational variable.

The passage from the standard {\it micro-canonical/canonical ensembles} to the 
{\it configurational ensembles} introduced in ref.~\cite{LMPR} is schematically 
illustrated in the steps 1. to 4. outlined below (in the formulae that follow we 
generically indicate with $r$ the whole set of variables necessary to describe 
the degrees of freedom of the system).
\begin{enumerate}
\item{\mbox{Make the replacement}}
\begin{eqnarray}
\begin{array}{lll}
U(r) &\to & A(r)\nonumber\\
\omega_U(E)=\int dr \delta(E-U(r)) &\to & \omega_A(a)=\int dr \delta(a-A(r))\nonumber
\end{array}
\end{eqnarray}
\item From seeds at random temperatures (see Appendix) collect configurations 
according to a Metropolis test with {\it multi-canonical} weight
\begin{eqnarray}
\begin{array}{lll}
[\omega_U(E)]^{-1}\equiv e^{-S(E)}&\to & [\omega_A(a)]^{-1}\equiv e^{-Q(a)}\, ,\nonumber
\end{array}
\end{eqnarray}
obtaining the configurational distribution (which may or may not be 
further elaborated)
\begin{eqnarray}
\begin{array}{lll}
\tilde{P}_U(r) &\to & \tilde{P}_A(r)\nonumber
\end{array}
\end{eqnarray}
\item{\mbox{Determine the best configurational distribution, $P$, 
satisfying the constraint}}
\begin{eqnarray}
\begin{array}{lll}
\langle U\rangle =\int dr U(r) {P}_U(r;\bar\beta)= \bar E &\to & 
\langle A\rangle =\int dr A(r) {P}_A(r;\bar\lambda)= \bar a\, ,  \label{CONSTR}
\end{array}
\end{eqnarray}
{\mbox{using the CMEM, which yields}}
\begin{eqnarray}
\begin{array}{lll}
{P}_U(r;\bar\beta)=\frac{1}{Z_U(\bar\beta)}\tilde{P}_U(r)e^{-\bar\beta U(r)}&\to & 
{P}_A(r;\bar\lambda)=\frac{1}{Z_A(\bar\lambda)}\tilde{P}_A(r)e^{-\bar\lambda A(r)}\nonumber\\
Z_U(\bar\beta)=\int dr \tilde{P}_U(r)e^{-\bar\beta U(r)} &\to & 
Z_A(\bar\lambda)=\int dr \tilde{P}_A(r)e^{-\bar\lambda A(r)} \nonumber
\end{array}
\end{eqnarray}
{\mbox{with the Lagrange multiplier implicitly fixed by~(\ref{CONSTR})}}
\begin{eqnarray}
\begin{array}{lll}
\bar\beta=\beta(\bar E) &\to & \bar\lambda=\lambda(\bar a)\, . \nonumber
\end{array}
\end{eqnarray}
\item The expectation value of $F=F(r)$ 
is computed by means of the re-weighting formula 
\begin{eqnarray}
\begin{array}{lll}
\langle F\rangle =\frac{\sum_{i=1}^{N_{conf}}[\omega_U(E_i)]^{-1} e^{-\bar\beta E_i}F_i}
{\sum_{i=1}^{N_{conf}}[\omega_U(E_i)]^{-1} e^{-\bar\beta E_i}}&\to & 
\langle F\rangle =\frac{\sum_{i=1}^{N_{conf}}[\omega_A(a_i)]^{-1} e^{-\bar\lambda a_i}F_i}
{\sum_{i=1}^{N_{conf}}[\omega_A(a_i)]^{-1} e^{-\bar\lambda a_i}}\, , \label{EXV}
\end{array}
\end{eqnarray}
{\mbox{where}}
\begin{eqnarray}
\begin{array}{lll}
E_i=U(r_i)\, ,\quad F_i=F(r_i) &\to & a_i=A(r_i)\, ,\quad F_i=F(r_i)\, ,\nonumber
\end{array}
\end{eqnarray}
and $N_{conf}$ is the number of collected configurations.
\end{enumerate}
Roughly speaking we may say that the computational strategy displayed 
in the left column is fine for energy related quantities, but not so 
much for structural quantities. On the contrary the new strategy 
outlined in the right column is expected to work appreciably well for 
structural quantities, but not as well for energy related quantities. 
To appropriately deal with them temperature must be brought back on stage.

\subsection{Introducing temperature}
\label{sec:IT}

Introducing the notion of temperature for a complex (fully flexible) system, 
like a polymer, is a delicate issue, because of the observation we already made 
that configurations only slightly different in their atomic spatial arrangement 
may have largely different (potential) energies. Consequently, as it turns out, 
it becomes more and more difficult to get the correct (Boltzmannian) energy 
distribution of the total available energy among the many degrees of freedom 
of the system as the temperature increases (despite the fact that at high 
temperature overcoming energy barriers may become easier). 

Actually, in the scheme we have just discussed there is room for the introduction 
of a sensible notion of temperature. This is done in two separate, but complementary 
steps. Temperature can be injected in the configurational probability distribution, 
$P_A$, if we know how the expectation values of the configurational variable we 
have chosen to fix ({\it i.e.}\ $\bar a$) depend on $T$. This dependence will 
in turn induce a $T$ dependence in the values of the Lagrange multipliers that 
are obtained by solving the constraint equation~(\ref{CONSTR}). 
Through eq.~(\ref{EXV}) this dependence is then passed to the expectation value of
any other configurational quantity one wishes to compute. It is important to remark 
that in a similar way dependence upon other environmental parameters can be 
introduced in the study of the physico-chemical properties of the system.

The temperature dependence induced through the method described above is not enough, 
however, to produce the correct $T$ behaviour of quantities that require an accurate 
thermalization of all the degrees of freedom of the system for their calculation. 
Examples of such quantities are the moments of the (potential) energy distribution. 
In these cases a local, extra thermalization step has to be carried out. This can be 
accomplished in the following way. Starting from each one of the recorded configurations, 
one performs a number of hybrid MC steps with velocities extracted from a 
Maxwell--Boltzmann distribution at the desired temperature. At the end of each 
MD block of moves configurations are subjected to a standard Metropolis test with 
acceptance/rejection probability given by $exp(-\beta H)$, where $H$ is the total 
(kinetic plus potential) energy of the system. In this way configurations 
are smoothly thermalized at the desired temperature and can be used to compute 
the {\it ensemble} averages~(\ref{EXV}). 

\subsection{An application to oligopeptides}
\label{sec:ANTO}

A first step in the study of protein folding properties can be the 
determination of the local propensity of the amino-acid chain to form 
$\alpha$-helix, $\beta$-sheet or other more or less structured arrangements.
\begin{figure}
{\begin{center}
\includegraphics[width=9.1cm]{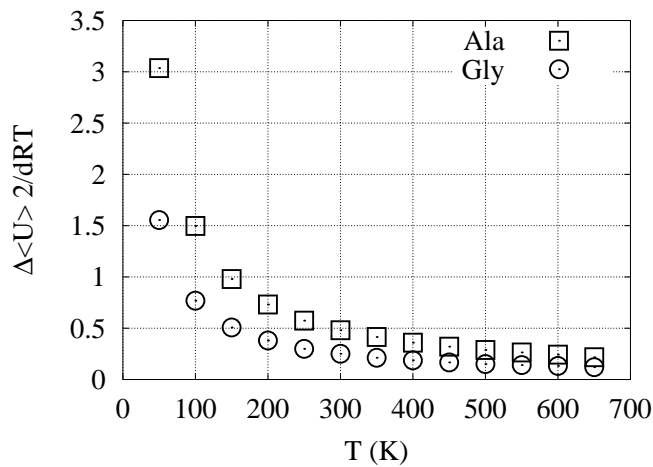}
\end{center}}
{\begin{flushleft}
\caption{Difference of average potential energy between disordered 
($\bar n_\alpha=0$) and ordered ($\bar n_\alpha=12$) states for Ala$_{12}$ (squares) 
and Gly$_{12}$ (circles) as function of temperature.}
\end{flushleft}}
\end{figure}
As an application of the considerations previously illustrated in this section,
I would like to briefly report on the interesting example considered in 
ref.~\cite{LMPR}, where the propensity to form $\alpha$-helix structures 
of two simple oligopeptides, {\it viz.}\ Gly$_{12}$ (a chain formed by 12 Glycine 
amino-acids) and Ala$_{12}$ (a chain formed by 12 Alanine amino-acids), was 
studied. The result of comparing data from the simulations of the two 
oligopeptides was that the order (folded) to disorder (unfolded) critical 
temperature is lower for Gly$_{12}$ than for Ala$_{12}$, implying that the 
propensity of Gly$_{12}$ to form $\alpha$-helix is lower than for Ala$_{12}$. 
The conclusion $T_c({{\mbox{Gly}}_{12}})<T_c({{\mbox{Ala}}_{12}})$, which is 
in agreement with experimental evidence, follows from Fig.~1 by 
identifying $T_c$ as the temperature at which the energy difference between 
the ordered and the disordered phase equals the equipartion energy. 
For completeness I give in the Appendix some detail on how the data of 
Fig.~1 are obtained following the strategy described in sect.~\ref{sec:ANP}.

I wish to end this section with two remarks. First of all the result 
one gets is rather robust and does not depend on the precise definition 
of critical temperature one decides to employ, as Ala$_{12}$ data all 
lie higher than Gly$_{12}$ data. Secondly and, most importantly, 
the possibility of determining such a non-trivial physico-chemical property of 
these systems should be regarded as a methodologically rather remarkable result 
because it is based on a first principle computation. It only relies, in fact, 
on the topology of the systems and the detailed properties of the atomistic model 
we took to describe the force field of the two oligopeptides. 

\section{Conclusions}
\label{CONCL}

It was my intention in this talk to convey to you the idea that the 
enormous computational power we have today at our disposal has been a 
crucial ingredient for the spectacular advances we have been witnessing 
in many research areas as well as in every-day-life applications and 
technological developments. I tried to do so by showing in a few  
significative examples, belonging to my field of competence, how 
the cultural layout upon which new ideas and methods have emerged 
have been appreciably influenced by the easy access to large-scale 
computing facilities and vice-versa.

In recent years a sort of mini-cultural revolution has been taking place.
The radically reductionistic paradigm, that so successful had proved to be 
in our quest for the fundamental laws of micro-physics, is appearing not 
to be fully adequate to deal with the challenge posed by the physics of, say, 
dynamical (non-linear) systems or the conceptual problems of modeling biological 
objects. Chances are that in these emerging fields of investigation notions like 
chaos or complexity will be going to play a central role. These notions have 
rapidly evolved from the initial physico-mathematical frameworks where they 
have been first introduced (non-linear dynamics and the physics of disordered 
systems). They have grown to the status of conceptual interpretative schemes 
under the stimulating pressure of the many successes they have led to in 
difficult numerical problems and the beneficial effect of the vast 
diversification of their field of application. 

\section*{Outlook}
\label{sec:OUT}

Let me conclude with a personal note. I started my career as a student of
Bruno Touschek who, as you certainly know, was the inventor of $e^+-e^-$ colliders
and the leader of the group of scientists that in Frascati built the first 
working storage ring, christened AdA, for ``Anello di Accumulazione''. A large 
fraction of theoreticians and students in Rome and Frascati were at that time (from 1966 
to 1970) busy with computing cross-sections for all sorts of $e^+-e^-$ processes. 
Computers were not based on transistors or chips, but on electronics tubes, 
and people were still busy with punching cards. So you were expected to carry on 
your calculations as much analytically as possible and only at the very end come 
up with some sensible approximation that one could work out numerically. Nevertheless 
Touschek was fond of electronic computers. He had clear in his mind their 
enormous potential in many strategic applications and he immediately suggested 
that computers should be used to simulate statistical systems with the idea 
of checking theory against actual numerical data. 

To a large extent this the same competent and inspired vision I found in Adriano's 
attitude towards research. One can identify a clear and consistent line of development 
in Adriano's scientific activity which, starting from his innovative studies on 
the role of topology in gluon-dynamics at finite and vanishing temperature, has 
naturally brought him in his more recent papers to attack the most difficult 
problem of all in QCD, the problem of understanding the mechanism underlying 
colour confinement (for a recent review see, for instance, ref.~\cite{CONF}).

I'm pretty sure that for many years to come Adriano's enthusiasm for research 
will still be a stimulating example for all of us: his ideas and intellectual 
ingenuity have had an enormous impact in lattice QCD, and more broadly in the 
whole field of high energy physics. 

Collaborating with Adriano was a privilege for me which has strongly 
influenced my approach to physics and shaped my scientific 
interests in research. I wish to thank him for that, but most of all 
for his invaluable and sincere friendship.

\section*{Acknowledgments}
It is a pleasure for me to thank the Organizers of the Fest for their 
kind invitation.

\section*{Appendix}
\label{sec:APP}

The definition of ordered and disordered thermodynamical 
states is given within the context of the strategy described in 
sect.~\ref{sec:SPP} by going through the following steps.

1) Configurations (seeds) are initially generated by sequential MD 
moves of fixed length by taking as starting system coordinates 
the coordinates of the last stored configuration and, as 
starting particle velocities, vector components extracted 
from a Maxwell--Boltzmann distribution at a random 
temperature, uniformly chosen at each MD step within zero
and a high-temperature limit (1000~K in the case at hand). This 
procedure can be proved to obey the detailed balance principle
and generates a time independent (stationary) conditional
probability, $P_c$. Although unknown, $P_c$ is perfectly
well defined and gives rise to an acceptable probability
distribution, $\widetilde{\cal P}^{(0)}$.

2) As a configurational variable relevant for the problem one is studying  
the average molecular $\alpha$-helicity, $N_\alpha$, is naturally 
taken. $N_\alpha$ is defined as the number of amino-acids with the 
two dihedral angles, C($i-1$)-N($i$)-C$\alpha$($i$)-C($i$) ($\phi_i$) and
N($i$)-C$\alpha$($i$)-C($i$)-N($i+1$) ($\psi_i$) within appropriately 
chosen bounds, which were taken to be  
$260^\circ\leq\phi_i\leq 320^\circ$, $i=2,\ldots,12$, 
and $293^\circ\leq\psi_i\leq 353^\circ$, $i=1,\ldots,11$.

3) The initial probability distribution, $\widetilde{\cal P}^{(0)}$, 
is improved by {\it multi-canonical} iterations leading from 
$\widetilde{\cal P}^{(k)}$ to $\widetilde{\cal P}^{(k+1)}$ by generating 
configurations that are accepted or rejected according 
to a Metropolis test based on the current $\alpha$-helicity number of 
states of the system, $\omega^{(k)}_{N_\alpha}(n_\alpha)$. 
The iterative procedure is stopped when some stability criterion 
is fulfilled and the last probability distribution, 
$\widetilde{\cal P}_{N_\alpha}$, is recorded.

4) For each oligopeptide the configurational probability distributions  
corresponding to the ordered and disordered phases are constructed 
from $\widetilde{\cal P}_{N_\alpha}$ by the CMEM, imposing the constraint 
$\bar n_\alpha=12$ or $\bar n_\alpha=0$, respectively.

5) At this point the two resulting probability distributions, 
${\cal P}_{N_\alpha}(\lambda({\bar n_\alpha=12}))$ and 
${\cal P}_{N_\alpha}(\lambda({\bar n_\alpha=0}))$, are 
thermalized at a set of temperatures ranging from $50$~K to $650$~K, in steps 
of $50$~K. From the latter the data points of Fig.~1 are obtained.


\section*{References}
\begin{thebibliography}{99}

\bibitem{APE}
P. Bacilieri {\it el al.}, ``{\it The APE project: a gigaflop parallel processor 
for lattice calculations}'', University of Rome ``{\it La Sapienza}'' 
preprint n.839 (1984), CERN-TH-4283-85  and ROM2F-85-28 preprint.

\bibitem{COL}
N.H. Christ, ``{\it The Columbia project: physics results and present status}'', 
Irvine Conference and Seillac Symposium (1987).

\bibitem{WIL}
K.G. Wilson, Phys. Rev. {\bf D10} (1974) 2445.

\bibitem{METR}
N. Metropolis, A.W. Rosenbluth, M.N. Rosenbluth, A.H. Teller and E. Teller,
J. Chem. Phys. {\bf 21} (1953) 1087.

\bibitem{PAR}
G. Parisi, ``{\it Statistical Field Theory}'', Frontiers in Physics vol. 66, 
Addison-Wesley (1988).

\bibitem{AMIT}
D. Amit, ``{\it Field Theory, the Renormalization Group, and Critical 
Phenomena}'', World Scientific (1984).

\bibitem{BERG}
B.A. Berg and T. Neuhaus, Phys. Lett. {\bf B267} (1991) 249:\\
B.A. Berg, Fields Institute Communications {\bf 26} (2000) 1 and cond-mat/9909236.

\bibitem{OKA}
A. Mitsutake, Y. Sugita and Y. Okamoto, 
Biopolymers (Peptide Science) {\bf 60} (2001) 96 and references quoted therein.

\bibitem{PMV}
M. M\'{e}zard, G. Parisi and M. Virasoro, ``{\it Spin Glass Theory and Beyond}'', 
World Scientific Lectures Notes in Physics vol. 9, World Scientific (1987).

\bibitem{MAR}
J.N. Banavar, A. Maritan, C. Micheletti and F. Seno, cond-mat/0105209 
and the many references quoted therein.

\bibitem{PJM}
G. Iori, E. Marinari and G. Parisi, J. Phys. A (Math Gen.) {\bf 24} 
(1991) 5349.

\bibitem{NP}
M.R. Garey and D.S. Johnson, ``{\it Computers and Intractability:
a Guide to the Theory of NP-Completeness}'', Freeman (New York, USA, 1979).

\bibitem{MON}
R. Monasson, Phys. Rev. Lett. {\bf 75} (1995) 2874.

\bibitem{ZEC}
M. M\'{e}zard and G. Parisi, Eur. Phys. J. {\bf B20} (2001) 217 
and cond-mat/0207121;\\
M. M\'{e}zard, G. Parisi and R. Zecchina, Science {\bf 297} (2002) 812;\\
M. M\'{e}zard and R. Zecchina, Phys. Rev. {\bf E66} (2002) 056126, cond-mat/0207194.

\bibitem{KSAT}
See, for instance, Satisfiability Library (www.satlib.org).

\bibitem{SCNAT}
The International Human Genome Mapping Consortium, ``{\it A physical map of 
the human genome}'', 
Nature {\bf 409}, 934-941 (Feb. 15, 2001);\\
The Celera Genomics Sequencing Team, ``{\it The sequence of the human genome}'', 
Science, {\bf 291}, 1304-1351  (Feb. 16, 2001).

\bibitem{MOO} 
G. Moore, "{\it Cramming more components onto integrated circuits}", 
Electronics Magazine (1965).

\bibitem{NN}
H.B. Nielsen and M. Ninomiya, Nucl. Phys. {\bf B185} (1981) 20 [Erratum {\it ibid.}
{\bf B195} (1982) 541]; Nucl. Phys. {\bf B193} (1981) 173 and Phys. Lett. {\bf B105} 
(1981) 219.

\bibitem{SVZ}
T. Banks, R. Horsley, H. R. Rubinstein and U. Wolff, Nucl. Phys. {\bf B190} (1981) 692;\\ 
M.A. Shifman, A.I. Veinshtein and V.I. Zakharov, Nucl. Phys. 
{\bf B147} (1979) 385, 448 and 519.

\bibitem{WV}
E. Witten, Nucl. Phys. {\bf B156} (1976) 269;\\
G. Veneziano, Nucl. Phys. {\bf B159} (1979) 213 and Phys. Lett. {\bf 95B} 
(1980) 90.

\bibitem{DGR}
A. Di Giacomo and G.C. Rossi, Phys. Lett. {\bf 100B} (1981) 481. 

\bibitem{RENO}
For a review, see: M. Beneke, Phys. Rep. {\bf 317} (1999) 1.

\bibitem{STOC}
F. Di Renzo, E. Onofri and G. Marchesini, Nucl. Phys. {\bf B457} (1995) 202;\\
G. Burgio, F. Di Renzo, G. Marchesini and E. Onofri, Phys. Lett. {\bf B422} (1998) 219;\\
F. Di Renzo and L. Scorzato, JHEP {\bf 10} (2001) 038.

\bibitem{RAK}
P.E.L. Rakow, hep-lat/0510046. 

\bibitem{LIMEU}
L. Li and Y. Meurice, hep-lat/0507034. 

\bibitem{CRE}
M. Creutz, Phys. Rev. {\bf D21} (1980) 2308 and BNL preprint 27 (1980) 995.

\bibitem{ACFP}
B. Alles, M. Campostrini, A. Feo and H. Panagopoulos, Phys. Lett. {\bf B324} (1994) 433. 

\bibitem{QTOP}
P. di Vecchia, K. Fabricius, G.C. Rossi and G. Veneziano, Nucl. Phys. 
{\bf B192} (1981) 392 and Phys. Lett. {\bf B108} (1982) 323;\\
K. Fabricius and G.C. Rossi, Phys. Lett. {\bf B127} (1983) 229.

\bibitem{DEF}
A. Di Giacomo, H. Panagopoulos and E. Vicari, Nucl. Phys. {\bf B338} (1990) 294.

\bibitem{COO}
M. Campostrini, A. Di Giacomo, H. Panagopoulos and E. Vicari, Nucl. Phys. {\bf B329} 
(1990) 683.

\bibitem{CHI} 
R. Narayanan and H. Neuberger,  Phys. Rev. Lett.  
{\bf 71} (1993) 3251 and   Nucl. Phys. {\bf B443} (1995) 305;\\   
P. Hasenfratz, Nucl. Phys. {\bf B} (Proc. Suppl.) {\bf 63} (1998) 53 and  
Nucl. Phys. {\bf B525} (1998) 401;\\   
H. Neuberger, Phys. Lett. {\bf B417} (1998) 141; {\it ibidem}  
{\bf B427} (1998) 353 and Phys. Rev. {\bf D57} (1998) 5417;\\   
P. Hern\'{a}ndez, K. Jansen, M. L\"{u}scher, Nucl. Phys. {\bf B552} (1999) 363. 
 
\bibitem{WALL}
D.B. Kaplan, Phys. Lett. {\bf B288} (1992) 342;\\  
Y. Shamir, Nucl. Phys. {\bf B406} (1993) 90;\\  
V. Furman and Y. Shamir, Nucl. Phys. {\bf B439} (1995) 54.

\bibitem{PERF}
P. Hasenfratz, V. Laliena and F. Niedermayer,  
Phys. Lett. {\bf B427}  (1998) 125.  

\bibitem{LU} 
M. L\"{u}scher, Phys. Lett. {\bf B428} (1998) 342.

\bibitem{GW}
P.H. Ginsparg and K.G. Wilson, Phys. Rev. {\bf D25} (1982) 2649.   

\bibitem{STAG}
J.B. Kogut and L. Susskind, Phys. Rev. {\bf D11} (1975) 395;\\
For a recent review, see C.T.H. Davies, hep-lat/0509046 and references therein.

\bibitem{GRTV}
L. Giusti, G.C. Rossi, M. Testa and G. Veneziano, Nucl. Phys.
{\bf B628} (2002) 234.

\bibitem{GG}
L. Del Debbio, L. Giusti and C. Pica, Phys. Rev. Lett. 
{\bf 94} (2005) 032003.

\bibitem{DIGNEW}
B. Alles, M. D'Elia, A. Di Giacomo and C. Pica, hep-lat/0509024.

\bibitem{TM} 
R. Frezzotti, P.A. Grassi, S. Sint and P. Weisz, Nucl. Phys. {\bf B} (Proc.
Suppl.) {\bf 83} (2000) 941 and JHEP {\bf 0108} (2001) 058;\\
R. Frezzotti, S. Sint and P. Weisz, JHEP {\bf 0107}
(2001) 048;\\
M. Della Morte, R. Frezzotti, J. Heitger and S. Sint,
JHEP {\bf 0110} (2001) 041;\\
R. Frezzotti and S. Sint, Nucl. Phys. {\bf B} (Proc. Suppl.) {\bf 106}
(2002) 814;\\
R. Frezzotti, Nucl. Phys. {\bf B} (Proc. Suppl.) {\bf 119} (2003) 140 and
{\bf 140} (2005) 134. 

\bibitem{FR1}
R. Frezzotti and G.C. Rossi, JHEP {\bf 0408} (2004) 007 
and Nucl. Phys. {\bf B} (Proc. Suppl.) {\bf 129} (2004) 880. 

\bibitem{FRC}
R. Frezzotti and G.C. Rossi,  Nucl. Phys. {\bf B} (Proc. Suppl.)
{\bf 128} (2004) 193.

\bibitem{FR2}
R. Frezzotti and G.C. Rossi, JHEP {\bf 0410} (2004) 070 and hep-lat/0509155.

\bibitem{FMPR}
R. Frezzotti, G. Martinelli, M. Papinutto and G.C. Rossi, hep-lat/0503034.

\bibitem{ENC}
K. Jansen, A. Shindler, C. Urbach and I. Wetzorke
[$\chi$LF Collaboration], Phys. Lett. {\bf B586} (2004) 432;\\
W. Bietenholz {\it et al.} [$\chi$LF Collaboration],
JHEP {\bf 0412} (2004) 044;\\
K. Jansen, M. Papinutto, A. Shindler, C. Urbach, I. Wetzorke,
[$\chi$LF Collaboration], hep-lat/0507010;\\
A.M. Abdel-Rehim, R. Lewis and R.M. Woloshyn, Phys. Rev. {\bf D71}
(2005) 094505 and Int. J. Mod. Phys. {\bf A20} (2005) 6159.

\bibitem{MONT}
F. Farchioni, R. Frezzotti, K. Jansen, I. Montvay, G.C. Rossi, E.E. Scholz,
A. Shindler, N. Ukita, C. Urbach, I. Werzorke, Eur. Phys. J. {\bf C39} (2005) 421;\\
F. Farchioni, K. Jansen, I. Montvay, E.E. Scholz, L. Scorzato, A. Shindler,
N. Ukita, C. Urbach and I. Wetzorke, Eur. Phys. J. {\bf C42} (2005) 73 and 
hep-lat/0512017;\\
F. Farchioni, K. Jansen, I. Montvay, E.E. Scholz, L. Scorzato,
A. Shindler, N. Ukita, C. Urbach, U. Wenger, I. Wetzorke, Phys. Lett.
{\bf B624} (2005) 324.

\bibitem{SW} 
B. Sheikholeslami and R. Wohlert, Nucl. Phys. {\bf B259} (1985) 572;\\
G. Heatlie, G. Martinelli, C. Pittori, G.C. Rossi and C.T. Sachrajda, 
Nucl. Phys. {\bf B352} (1992) 266.

\bibitem{OS}
K. \"Osterwalder and E. Seiler, Ann. of Phys. {\bf 110} (1978) 440.

\bibitem{BMMRT} 
M. Bochicchio, L. Maiani, G. Martinelli, G.C. Rossi and M. Testa,  
Nucl. Phys. {\bf B262} (1985) 331.

\bibitem{SYMA}
K. Symanzik, in ``{\it New Developments in Gauge Theories}'',
page 313, Eds. G. `t~Hooft {\it et al.}, Plenum (New York, USA, 1980);
``{\it Some topics in quantum field theory}'' in
``{\it Mathematical Problems in Theoretical Physics}'',  Eds. R.
Schrader {\it et al.}, Lectures Notes in Physics, Vol. 153, Springer (New
York, USA, 1982) and Nucl. Phys. {\bf B226} (1983) 187 and 205.

\bibitem{SHWUNEW}
S.R. Sharpe and J.M.S. Wu, Phys. Rev. {\bf D71} (2005) 074501 and 
Nucl. Phys. {\bf B} (Proc. Suppl.) {\bf 140} (2005) 323;\\
S.R. Sharpe, Phys. Rev. {\bf D72} (2005) 074510.

\bibitem{AB}
S. Aoki and O. B\"ar, Phys. Rev. {\bf D70} (2004) 116011 and hep-lat/0509002.

\bibitem{LUSS}
M. L\"{u}scher, S. Sint, R. Sommer and P. Weisz, Nucl. Phys. {\bf B478}
(1996) 365;\\
M. L\"{u}scher, S. Sint, R. Sommer, P. Weisz and U. Wolff, Nucl. Phys.
{\bf  B491} (1997) 323.

\bibitem{FGR}
R. Frezzotti, M. Golterman, M. Papinutto and G.C. Rossi, unpublished.

\bibitem{META1}
S. Aoki, Phys. Rev. {\bf D30} (1984) 2653 and 
Phys. Rev. Lett. {\bf 57} (1986) 3136.

\bibitem{META2}
S.R. Sharpe and R. Singleton, Jr., Phys. Rev. {\bf D58} (1998) 074501. 

\bibitem{CPT}
S.R. Sharpe and J.M.S. Wu, Phys. Rev. {\bf D70} (2004) 094029;\\ 
G. M\"unster, JHEP {\bf 0409} (2004) 035;\\ 
L. Scorzato, Eur. Phys. J. {\bf C37} (2004) 445. 

\bibitem{SHI}
A. Shindler, hep-lat/0511002;\\
F. Farchioni, P. Hofmann, G. M\"unster, K. Jansen, M. Papinutto, A. Shindler, 
U. Wenger, I. Wetzorke, I. Montvay, E.E. Scholz, N. Ukita, L. Scorzato and 
C. Urbach, hep-lat/0509131.

\bibitem{CHRMI}
K. Jansen {\it et al.} [$\chi$LF Collaboration], hep-lat/0507032 and hep-lat/0509036.

\bibitem{FRCY}
R. Frezzotti and G.C. Rossi, hep-lat/0511035.

\bibitem{ATFR}
M.P. Allen and D.J. Tildesley, ``{\it {Computer Simulation of Liquids}}", 
Clarendon Press (Oxford, UK, 1990);\\
D. Frenkel and B. Smit, ``{\it Understanding Molecular Simulations}'', Academic
Press (San Diego, USA, 1996);\\
U.H.E. Hansmann, ``{\it New Optimization Algorithms in Physics}'', 
VCH-Wiley (New York, USA, 2004), chap. {\it Protein Folding in Silico - 
The Quest for Better Algorithms}.

\bibitem{CP}
R. Car and M. Parrinello, Phys. Rev. Lett. {\bf 55} (1985) 2471;\\
A. Laio and M. Parrinello, PNAS {\bf 99} (2002) 12562.

\bibitem{CB}
H. Frauenfelder, K. Chu and R. Philipp, in ``{\it Biologically Inspired
Physics}", ed. L. Peliti, ed. Plenum Press (New York, 1991);\\
A. Chakrabartty and R.L. Baldwin, ``{\it Protein Folding: in Vivo 
and in Vitro}", eds. J. Cleland and J. King. American Physical Society 
(Washington, D.C., 1993).

\bibitem{PRU}
S.B. Prusiner, PNAS {\bf 95} (1998) 13363 and Science {\bf 278} (1997) 245.

\bibitem{AD}
D.J. Selkoe, Physiol. Rev. {\bf 81} (2001) 741;\\
P.M. Gorman and A. Chakrabartty, Biopolymers {\bf 60} (2002) 381.

\bibitem{FIB}
M.J. Welsh and A.F. Smit, Scientific American {\bf 273} (1995) 52;\\
M.A. Massiah {\it et al.}, Biochem. {\bf 38} (1999) 7453.

\bibitem{Branden99}
C. Branden and J. Tooze, ``{\it Introduction to protein structure}'', 
Garland Publishing Inc. (London, UK, 1999).

\bibitem{Saenger84}
W. Saenger, ed., ``{\it Principles of Nucleic Acid Structure}'', Springer-Verlag 
(New York, USA, 1984).

\bibitem{Rao98}
V.S.R. Rao, P.K. Qasba, P.V. Balaji and R. Chandrasekaran, 
``{\it Conformation of Carbohydrates}'', Harwood Academic Publishers (Amsterdam, 
The Netherland, 1998).

\bibitem{Mattice94} 
W.L. Mattice and U.W. Suter, eds., ``{\it Conformational Theory of Large Molecules. 
The Rotational Isomeric State Model in Macromolecular Systems}'', 
John Wiley \& Sons (New York, USA, 1994).

\bibitem{NOSEHOOVER}
S. Nos\`e, J. Chem. Phys. {\bf 81} (1984) 511;\\
W.G. Hoover, Phys. REv. {\bf A31} (1985) 1695;\\
S. Melchionna, G. Ciccotti and B.L. Holian, Mol. Phys. {\bf 78} (1993) 533;\\
G.J. Martyna, D.J. Tobias and M.L. Klein, J. Chem. Phys. {\bf 101} (1994) 4177.

\bibitem{KLEIN}
T. Husslein, D.M. Newns, P.C. Pattnaik, Q. Zhong, P.B. Moore and 
M.L. Klein, J. Chem. Phys. {\bf 109} (1998) 2826;\\
G.R. Dieckmann, J.D. Lear, Q. Zhong, M.L. Klein, W.F.
DeGrado and K.A. Sharp, Biophys. J. {\bf 76} (1999) 618;\\
R.M. Venable, B.R. Brooks, R.W. Pastor, J. Chem. Phys.
{\bf 112} (2000) 4822;\\
G. La Penna, S. Letardi, V. Minicozzi, S. Morante, G.C. Rossi and G. Salina, 
Eur. Phys. J. {\bf E5} (2001) 259. 

\bibitem{Kirkpatrick83}
S. Kirkpatrick, C. Gelatt, Jr. and M. Vecchi, Science {\bf 220} (1983) 671.

\bibitem{Okamoto01}
Y. Okamoto, ``{\it Encyclopedia of Optimization}'', Kluwer Academic (Dordrecht, 
The Netherlands, 2001), vol. III, chap. {\it Monte Carlo Simulated Annealing in 
Protein Folding}, pp. 425-439.

\bibitem{Wenzel99}
W. Wenzel and K. Hamacker, Phys. Rev. Lett. {\bf 82} (1999) 3003.

\bibitem{Holland75} 
J. Holland, ``{\it Adaptation in Natural and Artificial Systems}'', 
The University of Michigan Press (Ann Arbour, USA, 1975).

\bibitem{MP}
E. Marinari and G. Parisi, Europhys. Lett. {\bf 19} (1992) 451.

\bibitem{MODSIM}
B.A. Berg, G. La Penna, V. Minicozzi, S. Morante and G.C. Rossi, MODSIM
2003 Conference Proceedings, vol. 4, chap. {\it Multi-canonical Algorithms for Folding 
Processes}, pp. 1967 - 1972 (http://mssanz.org.au/modsim03/modsim2003.html).

\bibitem{LMPR}
G. La Penna,  J. Chem. Phys. {\bf 119} (2003) 8162;\\
G. La Penna, S. Morante, A. Perico and G.C. Rossi, J. Chem. Phys. {\bf 121} (2004) 10725.

\bibitem{CONF}
A. Di Giacomo, hep-lat/0510065 and references therein.

\end{thebibliography}
\end{document}